\documentclass[aps,pre,showpacs,showkeys,amsmath,amsfonts,amssymb]{revtex4} 
\usepackage{psfig}
\usepackage[dvips]{graphicx}
\usepackage{isolatin1}

\newcommand{\beq}{\begin{equation}} 
\newcommand{\eeq}{\end{equation}}
\newcommand{\bea}{\begin{eqnarray}} 
\newcommand{\eea}{\end{eqnarray}} 
\newtheorem{theorem}{Theorem} 
\newtheorem{corollary}{Corollary} 
 
\input epsf 
 
\begin{document} 
 
\title{Topological conditions for discrete symmetry breaking and phase transitions} 
 
\author{Fabrizio Baroni} 
\email{baroni@fi.infn.it} 
\affiliation{Dipartimento di Fisica, Universit\`a di 
Firenze, via G.~Sansone 1, I-50019 Sesto Fiorentino (FI), Italy}  
 
\author{Lapo Casetti} 
\email{lapo.casetti@unifi.it} 
\altaffiliation[Also at: ]
{CSDC, Universit\`a di Firenze, and INFN, Sezione di Firenze, Italy} 
\affiliation{Dipartimento di Fisica, Universit\`a di 
Firenze, via G.~Sansone 1, I-50019 Sesto Fiorentino (FI), Italy}  

\date{\today} 
 
\begin{abstract} 
In the framework of a recently proposed topological approach to phase
transitions, some sufficient conditions ensuring the presence of
the spontaneous breaking of a $\mathbb{Z}_2$ symmetry and of a
symmetry-breaking phase transition are introduced and discussed. A very simple
model, which we refer to as the hypercubic model, 
is introduced and solved. The main purpose of this model is that 
of illustrating the content of the sufficient conditions, but it is
interesting also in itself due to its simplicity. Then some mean-field 
models already known in the literature are discussed in the light of the
sufficient conditions introduced here. 
\end{abstract} 
 
 
 
\maketitle 

\section{Introduction}
\label{intro}  
Phase transitions are very common in nature. They are sudden
changes of the macroscopic behavior of a natural system composed by many
interacting parts occurring while an external parameter is smoothly varied.
Phase transitions 
are an example of {\em emergent} behavior, i.e., of collective  properties
having no direct counterpart in the dynamics or structure of  individual atoms
\cite{Lebowitz}. The successful description of phase transitions starting from
the properties of the microscopic interactions\footnote{Here the term
``microscopic'' must be understood in a wide sense, i.e., it refers to the
interactions between the degrees of freedom entering the Hamiltonian of the
system. It might denote the truly microscopic interactions between fundamental
particles, atoms, or molecules, or it may refer only to the interactions
between relevant degrees of freedom, as in spin systems, as well as to any
interaction between the individual components of the system, which may not be
microscopic at all: in the studies of the phase transitions occurring in
systems interacting via $1/r$ potentials, the ``microscopic'' interactions
could be gravitational forces between stars or even galaxies.}  between the
components of the system is one of the major achievements of equilibrium
statistical mechanics. From a statistical-mechanical point of view, in the
canonical ensemble, describing a system at constant temperature $T$, a phase
transition occurs at special values of the temperature called transition
points,  where thermodynamic quantities like pressure, magnetization, or heat
capacity, are non-analytic functions of $T$; these points are the  boundaries
between different phases of the system. Starting from the celebrated solution
of the two-dimensional Ising model by Onsager \cite{Onsager}, these
singularities have been indeed found in many models, and later developments
like the renormalization group theory \cite{Goldenfeld} have considerably
deepened our knowledge of the properties of the transition points, at least in
the case of continuous transitions, or  critical phenomena.

Yet, the situation is not completely satisfactory. First, in the canonical
ensemble these singularities occur only in the rather artificial case of
infinite systems: following an early suggestion by Kramers
\cite{Cohen_Uhlenbeck}, Lee and Yang \cite{LeeYang} showed that the
thermodynamic limit $N \to \infty$ ($N$ is the number of degrees of freedom,
and the limit is taken at fixed density)  must be invoked to explain the
existence of true singularities in  the canonical partition function $Z(T)$ and
then in the thermodynamic functions  defined as derivatives of $Z(T)$. Since in
the last decades many examples of transitional phenomena in systems far from
the thermodynamic limit have been found (e.g., in nuclei, atomic clusters,
biopolymers), a description of phase transitions valid also for finite systems
would be desirable. Second, while necessary conditions for the presence of a
phase transition can be found (one example is the above-mentioned need of the
thermodynamic limit in the canonical ensemble), 
nothing general is known about sufficient conditions: no general procedure is
at hand to tell if a system where a phase transition is not ruled out
from the beginning does have or not such a transition without computing $Z$:
only for some particular systems or class of systems
one can devise {\em ad hoc} procedures.
This might indicate that our deep understanding of this phenomenon
is still incomplete.

These considerations motivate a study of the deep nature of phase transitions
which may also be based on alternative approaches. One of such approaches,
proposed in Ref.\ \cite{cccp} and developed later \cite{physrep}, is based on
simple concepts and tools drawn from differential geometry and topology.  The
main issue of this new approach is a {\em topological  hypothesis}, whose
content is that at their deepest level phase  transitions are due to a topology
change of suitable submanifolds of configuration space, those where the system
``lives'' as the number of its degrees of freedom becomes very large.   This
idea has been discussed and tested in many recent papers 
\cite{franzosi,xymf,phi4,CSCP,euler,epl2003,ribeiro,grinza,kastner,tesi_fabrizio,garanin,andronico}.   
Moreover, the topological hypothesis has been given a rigorous  background by a theorem
\cite{theorem} which states that, at least for systems with short-ranged
interactions and confining potentials, 
topology changes in configuration space submanifolds are a
\emph{necessary} condition for a phase transition.   However, the converse is
not true (there are topology changes which are {\em not} connected with a phase
transition \cite{euler}), and no \emph{sufficient} topological conditions have
been  obtained yet. The problem of finding the {\em sufficient} conditions for
a phase transition remains one of the fundamental open problems in this field:
an answer to this problem would also  make the topological approach an ideal
candidate to define phase transitions in finite systems, for topology changes
in the relevant  submanifolds of configuration space do occur in finite
systems,  so that a criterion to discriminate the ``good'' ones would make them
a natural extension of the concept of a phase transition to finite $N$ case.

The present paper aims at contributing to the search for topological sufficient
conditions by considering, instead of the general problem of a generic phase
transition, the more particular -- but still very general and important from a
physical point of view -- case of the spontaneous breaking of a discrete
symmetry. Although a phase transition is a far more general phenomenon, which
is linked in general with the breaking of ergodicity\cite{Goldenfeld,Palmer} 
and may
or may not be accompanied by the spontaneous breaking of a symmetry, many
interesting phase transitions do occur in nature via the breaking of a
symmetry. One of the most familiar cases is ferromagnetism: in the
ordered phase the (continuous) rotational $O(3)$ 
symmetry of Heisenberg magnets is
spontaneously broken.  As to discrete symmetries, the paradigmatic example is
the Ising model on a lattice, or, if one wants to consider continuous variables, 
the lattice
$\varphi^4$ model, where if $d \geq 2$ a ferromagnetic transition exists and is
accompanied by the breaking of the global $\mathbb{Z}_2$ symmetry of the
Hamiltonian, i.e., the symmetry under the simultaneous reversal of all the
variables (the two-valued spins $s_i$ for the Ising model or the continuous real
variables $\varphi_i$ for the lattice  $\varphi^4$ model).

In the following we will consider only the case of a $\mathbb{Z}_2$ symmetry,
even if we believe that it should be possible to extend our results to general
discrete symmetries. We shall also restrict ourselves 
to systems described by continuous
variables, because the topological approach can be defined only for these
systems, even if in some cases we may refer to Ising-like discrete spin systems
for illustrative purposes. As we will show in Sec.~\ref{condition}, under fairly
general assumptions it is possible to state a sufficient condition for the
presence of a $\mathbb{Z}_2$-symmetry breaking phase transition essentially in
terms of the topology of the equipotential hypersurfaces in configuration
space, provided some additional conditions on the behavior with $N$ are
satisfied. Before stating and discussing this results, in
Sec.~\ref{qualitative} we will discuss at a general level the problem of the
breaking of a $\mathbb{Z}_2$ symmetry and the basis of the topological approach.
Then, after having stated the above mentioned result, in Sec.~\ref{models} we
will illustrate it introducing 
a simple abstract model, and in Sec.~\ref{phys_models}
we will discuss them in the light of some physical models already studied in the
literature. We will end with some
concluding remarks in Sec.~\ref{conclusions}.  

\section{General picture}
\label{qualitative}

\subsection{Phase transitions with $\mathbb{Z}_2$ symmetry breaking}
\label{qualitative_pt}
At a qualitative level, the physical mechanism underlying the spontaneous
breaking of a $\mathbb{Z}_2$ symmetry is quite well understood \cite{Parisi}. 
To begin with,
let us consider a simple example with a single degree of freedom: a particle of
unit mass in
a double-well potential $V(q)$, such that $V(q) = V(-q)$, 
at a fixed temperature $T$. The dynamics of the
particle will be described by a Langevin equation 
\beq
\ddot q = -\gamma \dot q - \frac{dV}{dq} +\eta (t)\, ,
\eeq
where $\eta (t)$ is a $\delta$-correlated white noise whose amplitude is related
to the friction coefficient $\gamma$ and to the temperature 
by the fluctuation-dissipation theorem,
\beq
\langle \eta(t) \eta(t+\tau) \rangle  = 2\gamma T \, 
\delta(\tau) ~.
\eeq
The ``magnetization'', i.e., the order parameter of the system, whose nonzero
value signals the breaking of the $\mathbb{Z}_2$ symmetry, is the time 
average of the position $q$. As long as the temperature is low, the dynamics 
of the particle is essentially
an activated process, with two widely separated time scales: one small scale
in which the particle oscillates in one of the two wells, and a large
time scale $\tau$ in which one observes jumps from one well to the other. On
time scales $t \ll \tau$, the symmetry appears to be broken, because the
particle is confined in one of the two wells and the finite-time order parameter
is nonzero:
\beq
\frac{1}{t} \int_0^t q(t') \, dt' \not = 0 ~~~~~~~ t \ll \tau~.
\eeq 
According to Kramers' theory \cite{Kramers}, 
\beq
\tau \propto \exp \left( \frac{\Delta E}{T} \right)~,
\label{tau}
\eeq
where we have set $k_B = 1$ as we shall always do from now on, 
$\Delta E$ is the height of the energy barrier the particle has to
overcome in order to jump from the minimum of 
one well to the other, and is a finite
quantity if $V$ is the potential energy of a single particle, so that, 
even if the timescale $\tau$ increases exponentially while decreasing
the temperature, the order parameter $\langle q \rangle$ 
vanishes for any finite temperature:
\beq
\langle q \rangle = \lim_{t\to\infty} \frac{1}{t} \int_0^t q(t') \, dt' = 0
~~~~~~~ \forall T > 0~,
\eeq
and symmetry breaking is possible only at $T = 0$. Note that 
increasing $\Delta E$ does not change the situation, unless one takes the limit
$\Delta E \to \infty$ where
the symmetry is broken for any value of $T$. In any case, no phase transition
between a symmetric and a broken-symmetry phase is allowed in this
single-particle system.

Nonetheless, we are interested in many-particle systems: it is just the number
of degrees of freedom, $N$, which plays a fundamental role to make a symmetry
breaking possible \cite{Parisi}. 
The potential energy $V(q_1,\ldots,q_N)$ is now a function
of $N$ variables, still $\mathbb{Z}_2$-symmetric, i.e., $V(q_1,\ldots,q_N) =
V(- q_1,\ldots, - q_N)$. The potential energy necessarily has two equivalent
absolute minima related by the symmetry, but multidimensionality  of
configuration space means that there could be many possible routes to go from
one minimum to the other. If now we denote by $\Delta E$ the {\em minimum}
barrier to jump in order to connect the two minima, for low enough $T$ 
Eq.~(\ref{tau}) still
holds\footnote{Equation (\protect\ref{tau}) can indeed be generalized to
a multidimensional situation \protect\cite{Hanggi}. 
The multiplicative factors neglected in Eq.~(\protect\ref{tau}), 
which in the single-particle case are essentially dimensional constants but 
now account for the entropic contribution, i.e.,
for the width of the wells and for the width of the saddle to cross, are more
complicated and important but this does not affect our reasoning which remains
at a qualitative level.}. 
This means that, if
$\Delta E$ grows with $N$, 
in the thermodynamic limit the equilibration time scale
$\tau$ becomes infinite and the system is trapped in one of the two wells even
for infinite times: the order parameter 
$\frac{1}{N}\langle \sum_{i=1}^N q_i \rangle$ is now
finite, 
\beq
\frac{1}{N}\left\langle \sum_{i=1}^N q_i \right\rangle = \lim_{t\to\infty} \lim_{N\to\infty}
\frac{1}{Nt} \int_0^t \sum_{i=1}^N q_i(t') \, dt' \not = 0
~~~~~~~ \forall\, T < T_c~,
\label{sb}
\eeq
and the symmetry is broken for finite temperatures below a critical temperature
$T_c$. Note that in Eq.~(\ref{sb}) the two limits 
{\em do not} commute, i.e., we must first
let the system go to the thermodynamic limit and then take the infinite-time
averages. Doing the other way round we would not get any transition, and no
symmetry breaking would be present \cite{Parisi}.

The reason why this happens only for $T<T_c$ and not for any $T$ is
due on the one hand to the fact that the separation of timescales, 
and thus the 
activated process picture, holds only for sufficiently low temperatures, and on
the other hand to that  
the above discussion is oversimplified: we have
neglected the role of entropy, and 
the high dimensionality of the configuration space ensures that for
sufficiently high $T$'s the entropy will always disorder the system. The above
argument could be made more stringent using free energy barriers instead of
energy barriers. However,
our purpose was only to show, using an intuitive dynamical argument, how in a
many-particle system symmetry breaking {\em can} occur in the thermodynamic
limit, not to prove that it {\em does} occur. 
We have thus seen that one of the basic ingredients 
for the possibility of symmetry
breaking is, besides the thermodynamic limit, 
that the height of the minimum barrier to overcome must grow with $N$: 
here is where, among other factors, the
dimensionality of the system comes in. To make a familiar example, 
in one-dimensional Ising systems $\Delta E$
is constant as $N$ grows, because after having
flipped one spin all the others can be flipped without any extra energy cost, 
while is proportional to
$\sqrt{N}$ in a two-dimensional system, because to flip a whole region of spins the
energy is paid at the perimeter of the region, whose length scales as $L^{d-1}$
for a $d$-dimensional lattice of length $L$. 
We recognize here the Landau-Peierls argument
\cite{statmech_book}
to prove the existence of a finite-temperature 
symmetry breaking in a two-dimensional Ising system,
and indeed the physical content of this argument is the same of the
``dynamical'' argument above, as to the energy part. The Landau-Peierls argument
is much more powerful because the entropic contribution, and then $T_c$, can be
estimated too, provided we can efficiently count the relevant configurations,
which however limits its applicability to Ising-like systems. The dynamical
argument is valid for general systems with continuous variables, 
but remains at a qualitative level and does not provide a clear sufficient
condition for the presence of a phase transition. 
As we shall see in the following, it is
possible to translate it into the topological language, which does allow to
state such a sufficient condition. But before doing that let us review the basis
of the topological approach.

\subsection{Basis of the topological approach}

As already mentioned in the Introduction, where also relevant references to
original papers were given, the topological approach 
to phase transitions is based on the ``topological hypothesis'' that
phase transitions are due to suitable topology changes in some submanifolds of
configuration space defined by the potential energy function. Here we want to
recall which is the basis of this approach. 

Let us consider a Hamiltonian system with $N$ degrees of freedom and standard
kinetic energy, described by the Hamiltonian
\beq 
{\cal H} = \frac{1}{2}\sum_{i=1}^N p_i^2 + V(q_1,\ldots,q_N)~, 
\label{ham_standard} 
\eeq 
where $V(q)$  
(from now on $q \equiv \{q \}_{i=1}^N$) 
is the potential 
energy and the $q_i$'s and the $p_i$'s ($i = 1,\ldots,N$) are,  
respectively, the canonical 
conjugate  coordinates and momenta, and are continuous variables; $q \in
M$, where $M$ is the $N$-dimensional configuration space manifold, and $V$ is
bounded below on $M$.
The configurational partition function of such a system
can be written as (we omit the $(h^N N!)^{-1}$ 
normalization factor because it is irrelevant for our discussion)
\beq
Z_N(\beta) = \int_0^\infty d(Nv) \, e^{-\beta Nv} \omega_N(v) 
\label{Z}
\eeq
where $\beta = T^{-1}$ and 
$\omega_N(v)$ is the density of states at (potential) energy per degree of
freedom $v$, i.e., 
the Liouville measure of the isopotential hypersurface $\Sigma_v$, 
\beq
\omega_N(v) = \mu(\Sigma_v) = 
\int_{\Sigma_v} \frac{d\Sigma}{\Vert \nabla V \Vert} ~,
\eeq
where 
$\Sigma_v$ is defined as
the $Nv$-level set of the potential energy $V(q)$,
\beq
\Sigma_v = \{ q \in M | V(q_1,\ldots,q_N) = Nv \}~,
\label{levelsets}
\eeq
and $d\Sigma$ is the volume element on
$\Sigma_v$. We write $\omega_N$ as  
\beq
\omega_N(v) = \left(a_N(v)\right)^{N}~,
\eeq
where 
$a_N(v) = \exp\left[s_N(v) \right]$ and $s_N(v)$ is the configurational entropy
per degree of freedom.
Now we can write (without loss of generality, we assume that the absolute 
minimum of $V$ is zero)
\beq
Z_N(\beta) = N \int_0^\infty dv \, e^{N (\log a_N(v) - \beta v}~, 
\eeq
and then,
as $N$ gets very large, we can evaluate the integral over $v$ in (\ref{Z}) by
replacing it with the largest value of the integrand:
\beq
Z_N(\beta) = \text{const.} \, e^{N \left[\sup_{v}(\log a_N(v) - \beta v\right]} 
\label{supremum}
\eeq 
where up to now we have only assumed extensivity, i.e.,
\beq
a(v) = \lim_{N\to\infty} a_N(v) = \left[ \mu(\Sigma_v)\right]^{1/N}
\label{a}
\eeq
exists and is finite, which, due to the physical meaning of the density of
states, amounts to requiring that the specific configurational 
entropy is well defined in the thermodynamic limit\footnote{This 
holds at least for systems with short-ranged interactions, but also in many systems where
interactions are extensive even if not additive like some mean-field models.};   
then we can write 
\beq
Z_N(\beta) \mathop{\longrightarrow}_{N\to\infty}  N e^{-N\beta
\overline{v}(\beta)}\mu\left(\Sigma_{\overline{v}(\beta)}\right) 
= N e^{-N\beta \overline{v}(\beta)}\int_{\Sigma_{\overline{v}(\beta)}} 
\frac{d\Sigma}{\Vert \nabla V \Vert} ~,
\eeq
showing that the only relevant contribution to the partition function comes from
a single isopotential hypersurface $\Sigma_{\overline{v}(\beta)}$, where
$\overline{v}(\beta)$ is the value of $v$ which realizes the supremum in
Eq.~\ref{supremum}; if we assume that the function $\beta v -  \log a(v)$ 
has a single minimum which
does not coincide with the extrema of the interval of definition,
$\overline{v}(\beta)$ is the solution of the saddle-point equation
\beq
\frac{d}{d\beta}\left[ \beta v - \log a(v)\right] = 0~,
\label{saddlepoint}
\eeq
and coincides with
the expectation value of $V/N$, 
\beq
\overline{v}(\beta) = \frac{1}{N}\langle V
\rangle = -\frac{1}{N}\frac{\partial \log Z_N(\beta)}{\partial \beta}~.
\label{average_v}
\eeq
This amounts to saying that as $N \to \infty$ 
the support of the equilibrium measure reduces to
the equipotential hypersurface $\Sigma_{\overline{v}(\beta)}$. 
In other terms, when computing the canonical ensemble average of a
function $A(q)$, we can write as $N$ gets very large
\beq
 \langle A \rangle (\beta) \mathop{\longrightarrow}_{N\to\infty}  
= \frac{1}{\tilde{Z}(\beta)}\int_{\Sigma_{\overline{v}(\beta)}} 
\frac{\left. A\right|_\Sigma 
d\Sigma}{\Vert \nabla V \Vert} ~,
\label{average}
\eeq
where $\left. A\right|_\Sigma$ is the restriction of the function
$A(q)$ to $\Sigma_{\overline{v}(\beta)}$ and 
\beq
\tilde{Z}(\beta) = \int_{\Sigma_{\overline{v}(\beta)}} 
\frac{d\Sigma}{\Vert \nabla V \Vert}~.
\eeq
A major topology change in a family of manifolds, depending on a continuous
parameter $v$, occurring at some $v_c$ may induce
singularities in the $v$-dependence of their volume, whence the
basic idea of the topological hypothesis: the deep origin of a phase transition 
might be concealed in the way the
configuration space is foliated in level sets of the potential energy function,
for a sufficiently ``strong'' topology change in the $\Sigma_v$'s or the $M_v$'s at some 
$v_c$ might induce a phase transition at a temperature such that $v_c =  \frac{1}{N} \langle V
\rangle$ because, as we have just seen, 
at very large $N$ the measure concentrates on a single
``slice''. 

An important remark is in order. All these results can be
reformulated considering the submanifolds $M_v = \{ q \in M | V(q) \leq Nv \}$
instead of the $\Sigma_v$'s;
the relation between these two families of submanifolds is $\Sigma_v = \partial
M_v$. The reason for the possibility of substituting the $M_v$'s to the
$\Sigma_v$'s is in the fact that the Liouville measure of $M_v$ is the same as
that of  $\Sigma_v$ when $N \to \infty$, and that topology changes of the
$\Sigma_v$'s do occur simultaneoulsy with those in the $M_v$'s apart from very
particular cases. In applications, using the $M_v$'s instead of the
$\Sigma_v$'s may be easier (for instance, in some cases 
Morse theory allows a
direct calculation of the topology changes in the $M_v$'s using the potential
energy as a Morse function \cite{euler,epl2003}).

As already noted in Sec.~\ref{intro}, the idea of the 
topological hypothesis has been discussed 
and tested in
some particular models, and a theorem has been proven showing that -- for a
wide class of systems -- topology changes in the $\Sigma_v$'s are a necessary
condition for a phase transition to occurr \cite{theorem}. 
In the following Section we are
going to show how a {\em sufficient} 
condition can be derived in the case of the
spontaneous breaking of a $\mathbb{Z}_2$ symmetry.

\section{Sufficient topological condition for $\mathbb{Z}_2$ symmetry breaking}
\label{condition}

Let us now consider a system of the class
(\ref{ham_standard}) with a potential energy which is $\mathbb{Z}_2$-invariant.
As we have shown in the previous Section, in the thermodynamic limit the
canonical ensemble average of a function of the coordinates 
is given by Eq.~(\ref{average}). Let us now consider, instead of a generic
function $A(q)$, a function whose average is a order parameter for the
$\mathbb{Z}_2$ symmetry breaking, i.e., 
\beq
A (q) = \frac{1}{N} \sum_{i=1}^N q_i ~.
\label{A}
\eeq
By inserting Eq.~(\ref{A}) into Eq.~(\ref{average}), at a first sight we are led
to conclude that no symmetry breaking is possible also in the thermodynamic
limit, because all the hypersurfaces $\Sigma_v$, being level sets of the
potential energy function $V$, must have all the symmetry of the function $V$
itself, and in particular the $\mathbb{Z}_2$ symmetry; hence\footnote{We are
implicitly assuming that also the integration 
measure $d\Sigma/\Vert\nabla V\Vert$ 
in Eq.~(\ref{average}) is still $\mathbb{Z}_2$-invariant in the
thermodynamic limit; this is perfectly reasonable because if $V(q)$ is
$\mathbb{Z}_2$-invariant, $\Vert\nabla V\Vert$ is $\mathbb{Z}_2$-invariant too. 
However, since it is not a uniform measure, there may be particular cases in
which it concentrates on submanifolds of $\Sigma_v$, and this may imply the
possibility of a symmetry breaking: we will discuss this case
in the following.} the order
parameter is zero for any value of $T$.

This conclusion is wrong. The $\Sigma_v$'s may be composed of two or more
disjoint connected components, and although the whole $\Sigma_v$ must respect
the invariance, each single connected component need not to be 
$\mathbb{Z}_2$-invariant: each one may be the image of another one under the
symmetry operation. If $\Sigma_v$ is made up of disjoint connected
components, then the definition (\ref{average}) of the ensemble average is
perfectly legitimate, but cannot be consistent with the actual behaviour of 
the system, because the representative point of the system can explore only one
of the connected components. This may be easily seen if we think of the
dynamics. Saying that when $N\to\infty$ the support of the measure is
$\Sigma_{\overline{v}(\beta)}$ is equivalent to say that the measure is formally
the standard Boltzmann weight
\beq
\varrho_\infty(q;\beta) = \frac{1}{Z_\infty} e^{-\beta V_\infty(q;\beta)} 
\eeq
with effective potential $V_\infty$ given by
\beq
V_\infty(q;\beta) = \left\{ 
\begin{array}{ccl}
Nv & \text{if} & q \in \Sigma_{\overline{v}(\beta)}\, ; \\
+\infty & \text{if} & q \in M \backslash \Sigma_{\overline{v}(\beta)}\, ;
\end{array}
\right.
\eeq
at $t=0$, the system will be in one of the disjoint components, 
and it will remain there forever: it can
never jump to another one because this would require to jump over an infinite
energy barrier. Hence, if $\Sigma_v = \Sigma_v^1 \cup \Sigma_v^2 \cup \cdots
\cup \Sigma_v^n$ with $\Sigma_v^a \cap \Sigma_v^b = 
\varnothing$ $\forall\, a,b$, the
restricted ensemble average being equal to a time average 
will be given by one of the following
\beq
 \langle A \rangle^a (\beta) \mathop{\longrightarrow}_{N\to\infty}  
= \frac{1}{\tilde{Z}^a(\beta)}\int_{\Sigma^a_{\overline{v}(\beta)}} 
\frac{\left. A\right|_{\Sigma^a} 
d\Sigma}{\Vert \nabla V \Vert} ~, ~~~~~~~~ a = 1,\ldots,n,
\label{average_correct}
\eeq
where the correct value of $a$ will be specified by the initial conditions, and 
\beq
\tilde{Z}^a(\beta) = \int_{\Sigma^a_{\overline{v}(\beta)}} 
\frac{d\Sigma}{\Vert \nabla V \Vert} ~, ~~~~~~~~ a = 1,\ldots,n~.
\label{Ztilde}
\eeq
If neither $\Sigma_v^a$ nor $A(q)$ are $\mathbb{Z}_2$-invariant, 
definitions (\ref{average}) and (\ref{average_correct}) do not yield the same
result; in particular, the average of $\frac{1}{N} \sum_{i=1}^N q_i$
according to Eq.~(\ref{average_correct}) may give a nonzero result,   
so that symmetry breaking is possible.

We note that the disjoint connected components $\Sigma_v^a$ of $\Sigma_v$ play
a role analogous to that of pure states in the standard approach to the formal
treatment of symmetry breaking \cite{Parisi,Ruelle},  while their union (i.e.,
the whole $\Sigma_v$),  plays the role of the mixed state.

We can now state the following
\begin{theorem}[sufficient topological condition for $\mathbb{Z}_2$ symmetry
breaking] \label{theorem1}
Let us consider a system of the class (\ref{ham_standard}) with $N$ degrees of
freedom and a potential
energy $V$ bounded below which is $\mathbb{Z}_2$-invariant. Let the entropy per
degree of freedom be well defined in the thermodynamic limit, i.e., the function
$a(v)$ defined in Eq.~(\ref{a}) exist and be continuous and piecewise
differentiable.
Let $\Sigma_v$ be the family of
equipotential hypersurfaces of the configuration space $M$ defined as in 
Eq.~(\ref{levelsets}). Without loss of generality, let $\text{min}(V)=0$.
Let $v'' > v' \geq 0$ be two values of the potential energy per degree of 
freedom $V/N$ such that $\Sigma_v  =  \cup_{a=1}^n \Sigma_v^a$ $\forall \, v \in
(v',v'')$, with $\Sigma_v^a \cap \Sigma_v^b = 
\varnothing$ $\forall\, a,b$, and such that $\forall a$ $\exists\, b \not = a \, 
: \mathcal{Z}(\Sigma_v^a) = \Sigma_v^b$ where $\mathcal{Z}$ is the
$\mathbb{Z}_2$-symmetry map on $M$, $\mathcal{Z}(q) = -q$.

Then, in the thermodynamic limit the $\mathbb{Z}_2$ symmetry 
is spontaneously broken for all the
temperatures $T\in (T',T'')$, where $T'' > T' \geq 0$.
\end{theorem}
{\bf Proof.} Thanks to the hypotheses on the function $a(v)$, 
the statistical average of
the potential energy per degree of freedom, $\overline{v}(T)$, 
is given by Eq.~(\ref{average_v}) and 
is monotonically increasing as $T$ is varied from
$0$ to $+\infty$, because $\frac{d\overline{v}}{dT} < 0$ would imply a negative
(configurational) 
heat capacity, which is forbidden in the canonical ensemble \cite{statmech_book}. 
Then there exists $T',T'' \geq 0$, 
$T' = \overline{v}^{-1}(v')$, $T'' = \overline{v}^{-1}(v'')$, 
such that $\overline{v}(T) \in (v',v'')$ $\forall \, T
\in (T',T'')$. For the sake of clarity let us assume that 
for $v \in (v',v'')$ the
equipotential hypersurface is made up only of two disjoint connected components,
$\Sigma_v  =  \Sigma_v^+ \cup  \Sigma_v^- $ $\forall \, v <
v'$, with $\Sigma_v^+ \cap \Sigma_v^- = 
\varnothing$ and $\mathcal{Z}(\Sigma_v^+) = \Sigma_v^-$; the extension to a larger
number of components is straightforward.
According to Eq.~(\ref{average_correct}), for $T \in (T',T'')$ 
the order parameter is  
\beq
 m^\pm = \left \langle \frac{1}{N} \sum_{i=1}^N q_i \right\rangle^\pm 
  \mathop{\longrightarrow}_{N\to\infty}  
= \frac{1}{N\tilde{Z}^\pm}\int_{\Sigma^\pm_{\overline{v}(\beta)}} 
\frac{ \sum_{i=1}^N q_i \,
d\Sigma}{\Vert \nabla V \Vert} ~,
\label{ordpar}
\eeq
and since the integrand is odd under ${\cal Z}$, $\tilde{Z}^+ =
\tilde{Z}^-$ -- this follows from the definition in 
Eq.~(\ref{Ztilde}) -- 
and $\mathcal{Z}(\Sigma_v^+) = \Sigma_v^-$ we have
\beq
m^+ = -m^- \not = 0~,
\eeq
because in order to have $m^\pm = 0$ each of the $\Sigma_v^\pm$'s should be
symmetric around $q = 0$, but this is impossible if they are disconnected and
$\mathbb{Z}_2$-invariant. Then the 
symmetry is spontaneously broken when $T \leq T'$. $\Box$

\begin{corollary}
\label{coroll}
Let us consider a system of the class (\ref{ham_standard}) with $N$ degrees of
freedom and a potential
energy $V$ bounded below which is $\mathbb{Z}_2$-invariant, an let it have two
degenerate distinct absolute minima $q^+= -q^-$ 
whose value is $\text{min}(V)=0$. Let the entropy per
degree of freedom exist as in Theorem \ref{theorem1}. Let 
Let $v' > 0$ be a value of the potential energy per degree of 
freedom $V/N$ such that all the $\Sigma_v$'s are homeomorphic for $v < v'$.

Then, in the thermodynamic limit the $\mathbb{Z}_2$ symmetry 
is spontaneously broken for all the
temperatures smaller than a finite value $T'$.
\end{corollary}
{\bf Proof.} When $v \to 0$, $\Sigma_v = \Sigma_v^+ \cup \Sigma_v^-$, where
$\Sigma_v^\pm$ are homeomorphic to $(N-1)$-spheres and 
$\Sigma_v^+ \cap \Sigma_v^- = \varnothing$. Since no topology change occurs until
$v'$, for all the temperatures smaller than $T' = \overline{v}^{-1}(v')$ we are
in the situation of Theorem \ref{theorem1}, whence the thesis. $\Box$

We note that the crucial ingredient to obtain the sufficient condition of
Theorem \ref{theorem1}
is that the value of the potential energy
below which the topology of the $\Sigma_v$'s is such that they are made up of
disjoint connected components must be proportional to $N$, and this agrees with
the qualitative reasoning of Sec.~\ref{qualitative_pt}. Being a sufficient
condition, but not a necessary one, this obviously does not rule out the 
possibility of having a symmetry breaking with less strong assumptions.
Moreover, we have proven that the symmetry is broken below $T''$, but neither
that it is restored above $T''$, nor that there is a phase transition at $T''$ 
or at any other $T$. 
To prove that, we have to strengthen the hypoteses, as in the
following
\begin{theorem}[sufficient condition for $\mathbb{Z}_2$-symmetry-breaking phase
transitions] \label{theorem2}
In addition to the hypotheses of Theorem \ref{theorem1}, let $v''' \geq v''$ be 
a value of the potential energy per degree of 
freedom $V/N$ such that $\Sigma_v$ is made of a single connected component for
$v > v'''$. Let
also the support of the canonical measure be the whole
$\Sigma_{\overline{v}(T)}$ when ${\overline{v}(T)} \geq v'''$ also in the
thermodynamic limit.

Then, in the thermodynamic limit there exist finite temperatures $T' <
T'' \leq T'''$ such that the $\mathbb{Z}_2$ symmetry 
is spontaneously broken for all 
$T \in (T',T'')$ and is restored for all $T \geq T'''$.
There is also at least a phase transition (or more than one) at $T_c$ such that 
$T'' \leq T_c \leq T'''$. If $v'' = v''' = v_c$, then there is just one phase
transition at $T_c = \overline{v}^{-1}(v_c)$. 
\end{theorem}
{\bf Proof.} Theorem \ref{theorem1} ensures that $m \not = 0$ for 
$T \in (T',T'') = (\overline{v}^{-1}(v'),\overline{v}^{-1}(v''))$.
Then, reasoning as in the proof of Theorem \ref{theorem1}, 
there exists $T''' > 0$, 
$T''' = \overline{v}^{-1}(v''')$ such that $\overline{v}(T) \geq v'''$ 
$\forall \, T \geq T'''$. Then, for $T \geq T'''$ the order parameter is  
\beq
 m = \left \langle \frac{1}{N} \sum_{i=1}^N q_i \right\rangle 
  \mathop{\longrightarrow}_{N\to\infty}  
= \frac{1}{N{Z}}\int_{\Sigma_{\overline{v}(\beta)}} 
\frac{ \sum_{i=1}^N q_i \,
d\Sigma}{\Vert \nabla V \Vert} = 0~,
\label{ordpar2}
\eeq
because the $\Sigma_{\overline{v}(\beta)}$'s are all symmetric: the symmetry is
restored. 
Hence, the order parameter $m$ 
is nonzero when $T \in (T',T'')$, and is constant and equal
to zero when $T > T'''$. Then the function $m(T)$ has at least a non-analytic
point for a temperature $T'' < T_c < T'''$, thus there is at least one
phase transition in the system. 
If $v''=v'''=v_c$, then there is a transition occurring at $T_c =
\overline{v}^{-1}(v_c)$. $\Box$

We note that the hypothesis that the support of the measure remains the whole 
$\Sigma_{\overline{v}(T)}$ when ${\overline{v}(T)} \geq v'''$ also in the
thermodynamic limit is essential, because without this assumption we cannot
prove that $m = 0$. If as $N\to \infty$ the support of the measure shrinks to a
submanifold of $\Sigma_{\overline{v}(T)}$, then the symmetry may remain broken
even if $\Sigma_{\overline{v}(T)}$ is $\mathbb{Z}_2$-symmetric. 
We believe that this case is not a purely academic one: 
it is probably what happens in at least one physically relevant 
example (the mean-field $\varphi^4$ model, see Sec.~\ref{phys_models}). 
This assumption was not necessary at all in proving Theorem \ref{theorem1},
because this may affect only the actual value of $m$, but not the fact that it
is nonzero.

These two theorems have rather strong assumptions, which may probably be
weakened. Moreover, for a generic many-particle system it is not an easy task at
all to characterize all the topology changes undergone by the $\Sigma_v$'s (see
e.g.~the discussion given in Refs.~\cite{physrep,euler,epl2003}), so that we are
not claiming to have derived a practical all-purpose method to prove the
existence of symmetry-breaking phase transitions in generic systems. 
Nonetheless, the theorems do allow one to 
make predictions, which are confirmed by the
analysis of some known models, to be discussed in Sec.~\ref{phys_models}. But
before doing that, let us discuss an abstract toy model we introduce in the
next Section, whose main purpose is to illustrate the theorems in a simple and
clear way, but which may have also some interest on its own.

\section{Hypercubic model}
\label{models}

We now introduce and solve an abstract model to enlighten, in a pedagogical
way, the content of the theorems proven in the last Section. This model is
rather abstract, and from a physical point of view it can be seen as a model of
a particle bouncing in a potential in an $N$-dimensional space, but we build
it starting directly with the equipotential hypersurfaces $\Sigma_v$.

\begin{figure}
\begin{center}
\includegraphics[width=6cm]{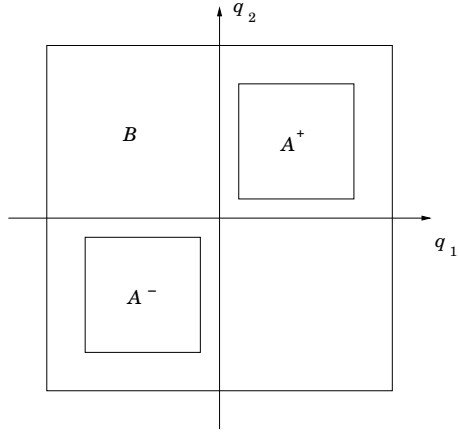}
\end{center}
\caption{Sketch of the hypercubes $A^\pm$ and $B$ for $N = 2$.}
\label{figure_hypercubes}
\end{figure}

The simplest $\mathbb{Z}_2$-invariant potential is a double square well in one
dimension, i.e., 
\beq
V(q) = \left\{ 
\begin{array}{ccl}
0 & \text{if} & a < |q| < b \, ; \\
v_0 & \text{if} & a > |q|  \, ; \\
+\infty & \text{if} & |q| > b \, ,
\end{array}
\label{squarewell}
\right.
\eeq
with $0 < a < b$. 
In this simple case the configuration space $M$ is just the real line
$\mathbb{R}$. Our first toy model, 
which we will refer to as the hypercubic model, 
is nothing but a generalization to $N$ dimensions of this double square well.
The configuraton space is now $\mathbb{R}^N$, and in $M$ we consider two
disjoint
hypercubes $A^+$ and $A^-$, symmetric under $\mathbb{Z}_2$, and a third
hypercube $B$, centered in the origin, such that $A^+,A^- \subset B$. Then we
define
\beq
V(q) = \left\{ 
\begin{array}{ccl}
0 & \text{if} & q \in A^\pm \, ; \\
Nv_c & \text{if} & q \in B \backslash \{A^+ \cup A^-\} \, ; \\
+\infty & \text{if} & q \in \mathbb{R}^N \backslash B \, .
\end{array}
\label{v_hypercubic}
\right.
\eeq
This potential is $\mathbb{Z}_2$-invariant by construction. The hypercubes
$A^\pm$
and $B$ are sketched in Fig.~\ref{figure_hypercubes} in the case $N = 2$.

The equipotential
hypersurfaces $\Sigma_v$ are then
\beq
\Sigma_v = \left\{ 
\begin{array}{ccl}
\varnothing & \text{if} & v < 0 ; \\
A^+ \cup  A^-  & \text{if} & v = 0; \\
\varnothing  & \text{if} & 0 < v < v_c \, ; \\
B \backslash \{A^+ \cup A^-\} & \text{if} & v = v_c; \\
\varnothing  & \text{if} & v > v_c \, .
\end{array}
\label{sigma_hypercubic}
\right.
\eeq
Then the topology of the $\Sigma_v$'s is such that
($\sim$ stands for ``is homeomorphic to'') 
\beq
\Sigma_v \sim \left\{ 
\begin{array}{ccl}
\varnothing & \text{if} & v < 0 ; \\
D^{N} + D^{N}  & \text{if} & v = 0 \, ; \\
\varnothing  & \text{if} & 0 < v < v_c \, ; \\
D_2^{N} & \text{if} & v = v_c \, ;\\
\varnothing  & \text{if} & v > v_c \, ,
\end{array}
\label{topo_hypercubic}
\right.
\eeq
where $D^N$ is a disk in $\mathbb{R}^N$, with $D_2^N$ we denote a two-punctuated
disk (a disk with two disjoint disks removed) and ``$+$'' stands for 
the disjoint union. 
The fact that apart from the values $v = 0$ and $v = v_c$ the $\Sigma_v$'s are
empty sets is due to the very singular nature of the potential, which has only
two possible values instead of a continuous interval: as $v$ increases
starting from values smaller than zero, the system potential energy
actually ``skips over'' all the values but $0$ and $v_c$, because these are the
only allowed values of $V/N$. Nonetheless, 
we see that the $\Sigma_v$'s of the hypercubic model undergo a topology change
as $v$ changes from $v=0$ to  $v = v_c$ of the kind described in Theorem \ref{theorem2}, in the
case $v' = v'' = v_c$. However, the situation is much clearer if we consider the
$M_v$ manifolds instead of the $\Sigma_v$'s: we have 
\beq
M_v = \left\{ 
\begin{array}{ccl}
\varnothing & \text{if} & v < 0 ; \\
A^+ \cup  A^-  & \text{if} & 0 \leq v < v_c \, ; \\
B & \text{if} & v  \geq v_c \, ,
\end{array}
\label{mv_hypercubic}
\right.
\eeq
and the the topology of the $M_v$'s is  
\beq
M_v \sim \left\{ 
\begin{array}{ccl}
\varnothing & \text{if} & v < 0 ; \\
D^{N} + D^{N}  & \text{if} & 0 \leq v < v_c \, ; \\
D^{N} & \text{if} & v  \geq v_c \, .
\end{array}
\label{topo_mv_hypercubic}
\right.
\eeq
Then the $M_v$'s of the hypercubic model undergo a topology change
as $v$ changes from $v=0$ to  $v = v_c$ precisely of the kind described in 
Theorem \ref{theorem2}, in the
case $v' = v'' = v_c$, 
and Theorem \ref{theorem2} (applied to the $M_v$'s) states that the hypercubic
model, in the thermodynamic limit $N \to \infty$, undergoes a phase transition
with $\mathbb{Z}_2$ symmetry breaking at a finite temperature $T_c$ such that 
$\frac{1}{N}\langle V \rangle(T_c) = v_c$. 
Let us see it explicitly, solving the model.

\begin{figure}
\begin{center}
\includegraphics[width=8cm,angle=-90]{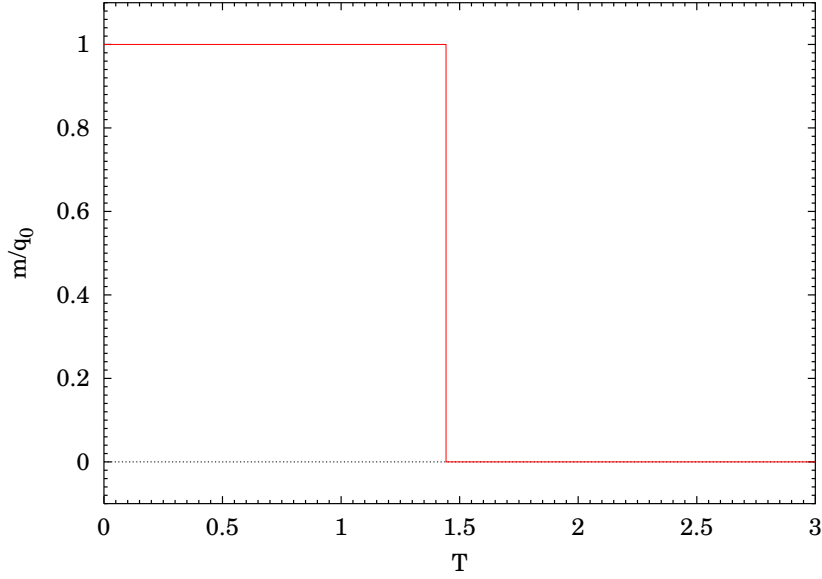}
\end{center}
\caption{Order parameter of the hypercubic model in the thermodynamic limit, as
a function of $T$. Here $v_c = 1$, $a = 1$, $b = 2a$, so that $T_c = (\log
2)^{-1}$. Only the positive branch is plotted.}
\label{figure_m_hypercubic}
\end{figure}

At any finite $N$, the configurational 
partition function of the hypercubic model
is 
\beq
Z_N(\beta) = \int_{\mathbb{R}^N} d^Nq \, e^{-\beta V(q)} = \int_{A^+} d^N q +
\int_{A^-} d^N q + e^{-\beta N v_c} \int_{B \backslash \{A^+ \cup A^- \}} d^N q 
\eeq
and denoting by $a$ and $b$ the length of the side of $A^\pm$ and $B$,
respectively, we obtain
\beq
Z_N(\beta) = 2a^N + (b^N - 2a^N) e^{-\beta N v_c}~,
\eeq
where  $b \geq 2a$ because $A^+,A^- \subset B$. Thermodynamic functions can then 
be computed at any finite $N$ and their limit as $N \to \infty$ can be studied
directly. We shall see that in the following, but before doing that we note that
when $N$ is large we can write
\beq
Z_N(\beta) \mathop{\longrightarrow}_{N\to\infty}  2a^N + b^N e^{-\beta N v_c} =
2 e^{N \log a} + e^{N(\log b - \beta v_c)}~,
\eeq
so that in the thermodynamic limit only the largest of the two exponentials
contributes to $Z$, and there will be a critical value $\beta_c$ of the 
inverse temperature $\beta$, given by the equation
\beq
\log a = \log b - \beta_c v_c 
\eeq
whose solution is
\beq
\beta_c = \frac{1}{v_c}\log\left(\frac{b}{a}\right)~,
\eeq
such that 
\beq
Z_N(\beta) \mathop{\longrightarrow}_{N\to\infty} \left\{ 
\begin{array}{ccl}
2 e^{N \log a} & \text{if} & \beta > \beta_c \, ; \\
 e^{N (\log b - \beta v_c)} & \text{if} & \beta < \beta_c \, .
\end{array}
\label{z_hypercubic}
\right.   
\eeq
This means that the system feels an effective double-well potential, with
potential energy zero and the
two wells separated by an infinite barrier, for $\beta > \beta_c$ and an
effective single-well, symmetric potential with potential energy $v_c$ 
when $\beta < \beta_c$. Hence the
symmetry is broken when $\beta > \beta_c$. The value of the order parameter
\beq
m = \frac{1}{N}\left\langle \sum_{i=1}^N q_i \right\rangle~,
\eeq   
where the average is the restricted one when the symmetry is broken, 
will be, in the thermodynamic limit,
\beq
m(\beta) = \left\{ 
\begin{array}{ccl}
\pm q_0 & \text{if} & \beta > \beta_c \, ; \\
0 & \text{if} & \beta < \beta_c \, ,
\end{array}
\label{m_hypercubic}
\right.   
\eeq
where $q_0$ is the value of all the coordinates $q_1,\ldots,q_N$ of the center
of the $A^+$ hypercube. The order parameter is plotted as a function of $T =
\beta^{-1}$ in Fig.~\ref{figure_m_hypercubic}. In the same limit, 
the average potential energy per degree of freedom will
be 
\beq
\langle v \rangle (\beta) = - \frac{1}{N} \frac{\partial}{\partial \beta} \log
Z_N(\beta) =  \left\{
\begin{array}{ccl}
0 & \text{if} & \beta > \beta_c \, ; \\
v_c & \text{if} & \beta < \beta_c \, ,
\end{array}
\label{avev_hypercubic}
\right.   
\eeq
so that the phase transition is a discontinuous (first-order) one. 

\begin{figure}
\begin{center}
\includegraphics[width=8cm,angle=-90]{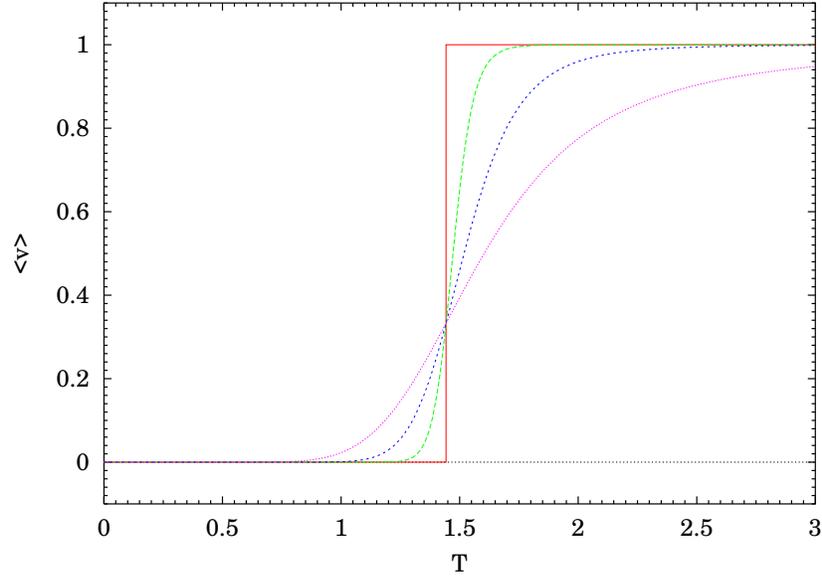}
\end{center}
\caption{Average potential energy of the hypercubic model as a function of $T$.
The different smooth curves
are the finite-$N$ result (\protect\ref{avev_hypercubic_N}) with $N = 10$, $20$
and $50$, while the piecewise constant curve is 
the $N \to \infty$ limit (\protect\ref{avev_hypercubic}).  Numerical values as in 
Fig.~\protect\ref{figure_m_hypercubic}.}
\label{figure_v_hypercubic}
\end{figure}

\begin{figure}
\begin{center}
\includegraphics[width=8cm,angle=-90]{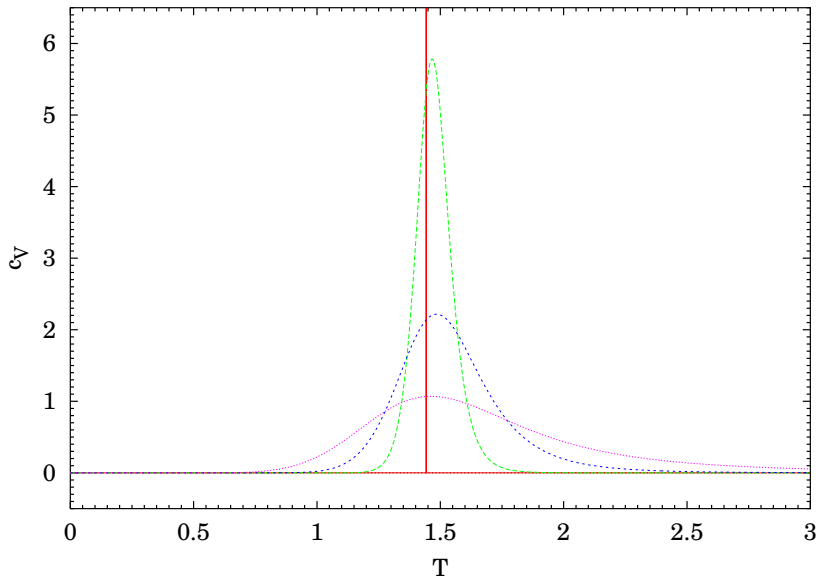}
\end{center}
\caption{As in Fig.~\protect\ref{figure_v_hypercubic} for the specific heat
$c_V$.}
\label{figure_cv_hypercubic}
\end{figure}

Computing the thermodynamic functions at finite $N$ one finds for the average
potential energy per degree of freedom
\beq
\langle v \rangle (\beta;N) = \frac{v_c (b^N - 2a^N) 
e^{-\beta N v_c}}{2 a^N + (b^N
- 2a^N) e^{-\beta N v_c}}~,
\label{avev_hypercubic_N}
\eeq
and for the configurational specific heat
\beq
c_V (\beta;N)  = \frac{2 N \beta^2 v_c^2 a^N (b^N - 2a^N) e^{-\beta N v_c}}
{\left[2 a^N + (b^N - 2a^N) e^{-\beta N v_c}\right]^2}~.
\label{cv_hypercubic_N}
\eeq
These functions are plotted in Figs.~\ref{figure_v_hypercubic} and
\ref{figure_cv_hypercubic}. We see that at any finite $N$ both $\langle v
\rangle$ and $c_V$ are regular functions, which converge (non uniformly) 
to the limiting non-analytic functions already determined before. As to 
the specific heat, since $\langle v \rangle$ is a piecewise constant function in
the thermodynamic limit, $c_V = 0$ everywhere but at $T_c$ where it has a
$\delta$-like singularity.

The hypercubic model is undoubtely very abstract, 
for it describes a particle in a
$N$-dimensional potential which is highly singular: yet it is 
extremely simple and neatly illustrates some
consequences of Theorem \ref{theorem2}. Moreover, it 
can be probably considered as one of
the most elementary models exhibiting a phase transition, so that it has a
pedagogical interest on its own.

However, as we shall see in the next Section, there are models, already known in the
literature, which can be analyzed in terms of the theorems of
Sec.~\ref{condition}, and whose potential energy function is indeed
regular and describes the interaction among microscopic degrees of freedom,
although in a not
completely realistic way due to its mean-field character. 

\section{Physical models}
\label{phys_models}

\subsection{Mean-field spherical model}

A physical example which can be exactly analyzed in terms of Theorem 
\ref{theorem2} is the mean-field spherical model, recently studied in
Ref.~\cite{ribeiro,kastner_inequiv}, which is the mean-field version of the
model originally introduced by Kac and Berlin in 1952 \cite{Kac}. 
It is an Ising-like system with continuous variables: 
the potential energy is
\beq
V(\varphi) = -\frac{1}{2N} \sum_{i,j=1}^N \varphi_i \varphi_j ~,
\label{v_spherical}
\eeq
where the variables $\varphi_i \in \mathbb{R}$, $i = 1,\ldots,N$ are
subject to the condition  
\beq
\sum_{i=1}^N \varphi^2_i = N~
\eeq
which constraints the variables to live on the $(N-1)$-dimensional sphere of
radius $\sqrt{N}$ centered in the origin of $\mathbb{R}^N$, 
whence the name ``spherical model''. The $\mathbb{Z}_2$ invariance of the
potential is apparent from Eq.~(\ref{v_spherical}). The potential energy is also
bounded, so that the potential energy per degree of freedom is bounded too: $v
\in [-\frac{1}{2},0]$.

The topology of the equipotential hypersurfaces $\Sigma_v$ of the mean-field
spherical model can be easily determined, and one finds
\cite{ribeiro,kastner_inequiv} that
\beq
\Sigma_v \sim \left\{ 
\begin{array}{ccl}
\varnothing & \text{if} & v < - \frac{1}{2} ; \\
\mathbb{S}^{N-2} + \mathbb{S}^{N-2}  & \text{if} & - \frac{1}{2} < v < 0 \, ; \\
\mathbb{S}^{N-2} & \text{if} & v = 0 \, .
\end{array}
\label{topo_spherical}
\right.
\eeq
We are in the situation described in Theorem \ref{theorem2}, which
then predicts that this model, in the canonical ensemble, has a phase transition
with $\mathbb{Z}_2$ symmetry breaking at a finite temperature $T_c$ such that
$\langle v \rangle (T_c)= 0$. This is indeed what happens, and the critical
temperature turns out to be $T_c = 1$ \cite{ribeiro}.

The mean-field spherical model is thus 
a nice illustration of the consequences of
Theorem \ref{theorem2}.

\subsection{Mean-field $\varphi^4$ model}

An example where Theorem \ref{theorem1} holds is provided by another mean-field
version of a continuous-spin Ising model, the mean-field $\varphi^4$ model
\cite{tesi_fabrizio,garanin,andronico}. The interaction potential is 
\beq 
V(\varphi) = \frac{J}{N} \sum_{i,j=1}^N \varphi_i
\varphi_j +  \sum_{i=1}^N \left( - \frac{1}{2} \varphi_i^2 +
\frac{1}{4}\varphi_i^4\right)~,   
\label{V_phi4mf} 
\eeq 
where $J>0$ is a coupling constant and $\varphi_i \in \mathbb{R}$, $i =
1,\ldots,N$. Again the $\mathbb{Z}_2$ symmetry is apparent from
Eq.~(\ref{V_phi4mf}); this potential is bounded below, but not above, and $v
\in [v_{\text{min}},+\infty)$, where $v_{\text{min}} = -\frac{(J + 1)^2}{4}$.
There are two equivalent distinct minima, $\varphi_i = \pm \sqrt{J + 1}$, $i =
1,\ldots,N$, then as $v \to v_{\text{min}}$
\beq
\Sigma_v \sim \mathbb{S}^{N-1} + \mathbb{S}^{N-1}~. 
\eeq 
It has been numerically 
shown \cite{tesi_fabrizio} that for any $N$ and any $J > 0$
there are no topology changes\footnote{More precisely, in
Ref.~\protect\cite{tesi_fabrizio} it has been shown that there are no topology
changes in the manifolds $M_v$ whose boundaries are the $\Sigma_v$'s, i.e.,
$\Sigma_v = \partial M_v$. Nonetheless, since $M_v \sim \mathbb{S}^{N} +
\mathbb{S}^{N}$ as $v < v'$ and $\partial \mathbb{S}^{N} = \mathbb{S}^{N-1}$, we
have $\Sigma_v \sim \mathbb{S}^{N-1} + \mathbb{S}^{N-1}$ as $v < v'$.} 
in the
$\Sigma_v$'s as long as $v < v'$, where $v' < 0$ 
is a finite value which grows when
$J$ grows. Then, this model fulfills the hypotheses of Theorem \ref{theorem1},
or more precisely those of Corollary \ref{coroll}, so that the $\mathbb{Z}_2$
symmetry must be broken below a finite temperature $T'= \overline{v}^{-1}(v')$.
This is precisely what
happens: the magnetization $m(T)$, which can be exactly calculated in the
thermodynamic limit due to the mean-field character of the model, is nonzero for
$T < T'$. For instance, as $J =
\frac{1}{2}$, $v' \simeq -0.35$ and $T' \simeq 0.3$, while the critical
temperature below which the symmetry is broken is $T_c \simeq 0.4$; when $J =
1$, $v' \simeq -0.25$ and $T' \simeq 0.9$, while $T_c \simeq 1$
\cite{tesi_fabrizio}.

However, as $v > 0$, it has also 
been shown \cite{tesi_fabrizio,garanin,andronico} that  
\beq
\Sigma_v \sim \mathbb{S}^{N-1} ~, 
\eeq 
but at least for sufficiently large values of $J$ the $\mathbb{Z}_2$ symmetry
remains broken also for temperatures $T > T'' = \overline{v}^{-1}(0)$. Then,
although this model is a clear illustration of Corollary \ref{coroll}, it 
does not behave as predicted  by Theorem \ref{theorem2}, at variance with the
mean-field spherical model. This means that at least one of the hypotheses of
Theorem \ref{theorem2} does not hold for the mean-field $\varphi^4$ model. As we
will discuss elsewhere \cite{tobepub}, in this case it is not true that the
support of the equilibrium measure remains the whole $\Sigma_{\overline{v}}$
even in the thermodynamic limit, due to the additional constraint provided by
the fact that the function whose average is the 
order parameter enters the potential\footnote{This is true also for the
spherical model and in other cases like the mean-field $XY$ model 
\protect\cite{euler}; nonetheless, in
the latter cases $V$ can be written as a function of the order parameter alone,
so that even if it may restrict the support of the measure to a subset of the
$\Sigma_v$, it cannot change its symmetry properties. Rigorosuly speaking, then,
also for the mean-field spherical model one should say that Theorem
\protect\ref{theorem2} holds even if its hypotheses may not be completely
fulfilled.}

\section{Concluding remarks}
\label{conclusions}

We have shown that the topological approach to phase transitions allows one to
translate an intuitive, qualitative picture of the origin of
discrete-symmetry-breaking phase transitions into a sufficient condition for
such phenomenon to occur. This is a first step towards obtaining more general
sufficient conditions for phase transitions, thus towards 
filling a gap in our present
understanding of these ubiquitous and fascinating phenomena. The topological
sufficient conditions have been derived here in the case of $\mathbb{Z}_2$
symmetry, but the proofs of Theorems \ref{theorem1} and \ref{theorem2} are
easily adaptable to more general discrete symmetries: we chose to restrict
ourselves to $\mathbb{Z}_2$ for the sake of simplicity, and because of the
great importance of this particular 
symmetry in the development of our understanding of 
the physics of phase transitions. We note that our sufficient conditions are not
exclusively topological in nature, i.e., they cannot be formulated in terms of
the topological properties of the $\Sigma_v's$ alone: for example, 
we have also to introduce
some hypotheses on the behavior with $N$ of the values of the energy at which
some topology changes must happen. Nonetheless, these additional requirements
appear absolutely natural, and do not change the fact that the basic ingredient
of the two theorems proven here comes from topology.

A simple model introduced here, the hypercubic model, and some mean-field models
already known in the literature have been discussed in the light of 
these two theorems. It would obviously be particularly interesting to
investigate some more realistic systems using these results. The natural candidate
for such an investigation is the $\varphi^4$ model with short-range interactions
in two or more dimensions. 
At present, only numerical results for the topology of the
$\Sigma_v$'s are available for this model
\cite{phi4} which do not yet allow  
to state wheteher the hypotheses of some of the two theorems are fulfilled for
this model. This is clearly one of the natural lines of development for future
investigations. 

We are aware that these sufficient conditions may well be not ``optimal'' at
all, in the sense that it may be
possible to weaken the hypotheses; moreover, it is possible that in some cases
the actual physical relevance of the particular phenomenon described here is
small, because other mechanisms may be at work \cite{Hahn}. Nonetheless, 
the availability of a sufficiency criterion, although restricted to a particular
class of transitions and maybe not optimal yet, 
opens also the possibility of using
the topological approach to define phase transitions in finite systems, because
the topology changes in the manifolds $\Sigma_v$ which are at the basis of phase
transitions do occur also in finite systems: work is in progress along this
line. Moreover, the phenomenology of ``disconnection borders'' found
in some classical spin systems (see e.g.\ Ref.~\cite{Borgonovi}) might be well
related to our present results in the perspective of studying transitional
phenomena in finite systems. 

In conclusion, we believe that the present work adds some new insight to the
topological approach to phase transitions, confirming its great potentialities.




\end{document}